\documentstyle[12pt]{article}
\newcommand{\beq} {\begin{equation}}
\newcommand{\eeq} {\end{equation}}
\begin{document}
\section{Introduction}

Discrete kinetic theory, and notably the
Lattice Boltzmann method has met with significant success in the
recent years for the numerical simulation of a wide host of fluid 
phenomena, ranging from multiphase flows in grossly irregular geometries, up to 
highly turbulent homogeneous incompressible flows \cite{PHD,BSV,JSP}.
Main drivers behind the method are ideal amenability to parallel computing,
easy handling of grossly irregular geometries and physical soundness.  

The Lattice Boltzmann Equation (LBE) 
is {\it more} than a plain Navier-Stokes solver; actually, it is a
{\it minimal hyperbolic superset} of the Navier-Stokes equations.
In fact, in the process taking from the kinetic to the hydrodynamic level,
along with standard primitive fluid fields such as density, speed and
temperature, there is enough information left to describe the dynamics of a 
genuinely hydrodynamic stress tensor as well as that of a passive scalar.
This property permits to extend the applicability of
LBE techniques beyond 'plain' hydrodynamics, with only minor modification
to the basic discrete equation.
To date, this feature has been exploited in various ways, to describe
thermal convection, magnetohydrodynamics (\cite {QSO} and references therein)
and also subgrid modeling \cite {SOM}.

All of the above refers to the 24-speed FCHC (Face Centered HyperCube) lattice
\cite{DHU}, the ancestor of modern Lattice Boltzmann schemes, introduced by D'Humieres,
Frisch and Lallemand back in 1987.
Recently, FCHC has been superseded by swifter Lattice BGK schemes
\cite{CHE1,QIAN} which can boast a simpler (diagonal) collision operator as well as
a significantly smaller number of discrete speeds (14 instead of 24).

In spite of these undeniable advantages, we shall argue that the LBE-FCHC scheme still
holds a certain interest for it provides the unique opportunity to integrate passive
and active scalars dynamics within the very same scheme tracking the flow variables,
with no need to extend the corresponding set of discrete speeds.

The scope of this paper is twodfold; first, to revisit
the symmetry properties of LBE-FCHC schemes, second to present
a novel potential application of these symmetries to the 
field of reactive flow dynamics.

\section{Reactive flow dynamics}

The set of hydro-chemical equations governing the dynamics
of a $S$-species reactive flow read as follows ($k,l$ run over the spatial dimension): 

\begin{eqnarray}
\partial_t \rho_s + \partial_{l} (\rho_s u_l) = 
\partial_l D_s \rho  \partial_l ( \rho_s / \rho ) + S_s
\;,\;\;\;\;\;s=1,S
\end{eqnarray}

\begin{eqnarray}
\partial_t \rho u_k + \partial_{l} (\rho u_k u_l) = 
\partial_l \nu \rho \partial_l u_k - \partial_k p  
\end{eqnarray}

\begin{eqnarray}
\partial_t \rho T + \partial_{k} (\rho u_{k} T) = 
\partial_l \chi \rho \partial_l T
+ \dot Q 
\end{eqnarray}

where $\rho_s$ is the density of the $s-th$ species, 
$\rho$ is the fluid density ($\rho = \sum_s \rho_s)$, $T$ the fluid temperature, 
$p$ the fluid pressure, $u_k$ the flow field, $D$, $\nu$ and $\chi$ the 
mass, momentum and energy kinetic diffusivity of the fluid.
Here $S_s$ is the mass density input to species $s$
from chemical reactions, and $\dot Q$ is the heat input rate as obtained
by summing all over the reactions: $\dot Q=\sum_r Q_r \dot \omega_r$
, where $\dot \omega_r$ is the progress rate (moles/s) of reaction $r$.

Combustion is governed by the competition between hydrodynamics
(diffusion and convection) and chemistry; the 
relative strength of this competion is measured by a dimensionless parameter
known as the Damkohler number, defined as $Da = \frac{\tau_h}{\tau_c}$
where $\tau_h$ and $\tau_c$ are typical hydrodynamic and chemical timescales respectively,

Low Damkohler's characterize quasi-inert flows while high Damkholer's 
are associated with fast ($\tau_c << \tau_h$) chemical reactions.

In this paper we shall be concerned with the 'fast' combustion limit, i.e.
reactive flows in which chemistry proceeds much faster than hydrodynamics
so that as soon as mixing is completed the fuel instantly burns 
("mixed-is-burned"). 

In this limit, combustion tends to localize on thin regions ("reaction sheets") whose
thickness becomes vanishingly small as the inverse Damkohler number goes to zero
\cite{WIL}.
This brings about a drastic simplification in the mathematical model
because the focus can be shifted onto the topology of the reaction
sheet. It can be shown that for the case of {\it non-premixed} flames, i.e.
flames where fuel and oxidizer are fed into the system in separate 
streams, and under the equidiffusional assumption
($D= \nu = \chi$), the topology of the reaction sheet is properly described by
a passive scalar, say $Z$, evolving according to the usual 
advection-diffusion equation:

\begin{equation}
 \rho D_t Z = \partial_k \rho D \partial_k Z
\end{equation}

where $D_t = \partial_t + u \partial_x +v \partial_y + w \partial_z $
is the fluid substantial derivative.

The underlying idea is that while species transform one into another
under reactive events, there are quantities 
(typically elemental mass fractions) which are unaffected by chemistry
and serve therefore as slow modes enslaving the system dynamics. 
As a result, in the limit where chemistry proceeds much faster than
hydrodynamics, all species concentrations are 'frozen-in' byproducts
of the conserved scalars. 
With reference to a diffusive non-premixed flame, it is readily checked that 
a convenient choice for $Z$ is provided by the fuel mixture fraction,
defined as the mass ratio of fuel in the unburned mixture
versus fuel mass ratio in the stream, 

\beq
 Z= Y_{fu}/Y_{fs}
\eeq

where $Y$ stands for mass fraction and the subscripts $u,s$ refer to the 
unburned and stream fuel respectively.

For a single-step irreversible reaction of the form

$Fuel + Oxidizer \rightarrow Products$

combustion takes place around the surface where reactants and products 
come in stoichiometric proportion. 
The stoichiometric surface is defined by the relation:

$Z(x,y,z) = Z_{st} = 1/(1+ A Y_{f,s} / Y_{o,s})$

where $A$ is the ratio of the fuel to oxidizer stoichiometric 
coefficients times the corresponding molecular weights \cite{PET}.
In the limit of infinitely fast chemistry, the stoichiometric surface
represents a front of discontinuity for the spatial derivatives of 
the hydrodynamic variables. Within this limit, and with the additional
assumption of steady state and adiabatic conditions, all quantities 
become a simple piecewise linear map of $Z$. 
With reference to {\it lean} ($0 < Z < Z_{st}$, superscript "L") and {\it rich} 
($Z_{st} < Z < 1$), superscript "R") portions of the mixture, these mappings read as follows:

\beq
Y^L_{fb} (Z) = 0     
\eeq

\beq
Y^R_{fb} (Z) = Y_{fs} (Z - Z_{st})/(1 - Z_{st})    
\eeq

\beq
Y^L_{ob} (Z) = Y_{os} (1 - Z/Z_{st})     
\eeq

\beq
Y^R_{ob} (Z) = 0    
\eeq

\beq
T^L_{b}(Z) = T_{in} + (Z/Z_{st}) (T_{st} - T_{in})   
\eeq

\beq
T^R_{b}(Z) = T_{in} + (\frac{1-Z}{1-Z_{st}}) (T_{st} - T_{in})
\eeq

where subscript "b" means "burned".

Here $T_{in}$ is the inlet stream temperature 
(we assume stream fuel and oxidizer at the same temperature) and
the actual value of $T_{st}$ depends on the specific chemical reaction.
Note that in the unburned portion of the mixture the dependence
on $Z$ disappears ($T_u = T_{in}$).

Within this approximation the heat source reduces to 
$
\dot Q = Q_r \rho \Delta Y_{f} / W_{f} \delta (Z-Z_{st})
$
where $Q_r$ is the reaction heat (Joule/mole) and
$\Delta Y_{f}$ represents the change of the fuel mass fraction
between the unburned ($Y_f = Z Y_{fs})$) and the burned ($Y_f=Y_{fb}$) state.
The presence of the Dirac's delta reflects the instantaneous nature
of the chemical reactions.

\section{Extended LBE for reactive flows}

The LBE is a minimal discrete Boltzmann equation reproducing
Navier-Stokes hydrodynamics in the limit of small Knudsen
numbers, i.e. particle mean free path much smaller than 
typical macroscopic variation scales \cite{BSV}.
We shall refer to the 24-speed, single-energy, four-dimensional FCHC 
(Face Centered HyperCube) lattice defined by the condition \cite{DHU}

$\sum_{k=1,4} c_{ik}^2 = 2 \;\;\; c_{ik}= (0, \pm 1 )$

The FCHC-LBE takes the form of a set of first-order finite difference
equations:

\beq
f_i (x_k + c_{ik}, t+1) - f_i (x_k,t) = \sum_{j=1}^{24} A_{ij}(f_j -f_j^e)
\eeq

where $f_i, i=1,24$ is a set of 24 populations moving along a corresponding set
of discrete speeds $c_{ik}$ and $k=1,4$ runs over the spatial dimensions. 

The eq.(12) is naturally interpreted as a multi-relaxation scheme in
which non-equilibrium gradients are brought back to the local
equilibrium $f_j^e$ by scattering events mediated by the matrix $A_{ij}$.
The local equilibrium populations are chosen in the form of a discretized Maxwellian
expanded to second order in the Mach number in order to retain 
convective effects.

\begin{equation}
f_i^e = \frac{\rho}{24} (1 + 2 c_{ik} u_k + 2 (c_{ik} c_{il} - \frac{1}{2} \delta_{kl}) u_k u_l)
\end{equation}

Under the constraint of point-wise mass and momentum conservations
($\sum_i A_{ij} = \sum_i c_{ik} A_{ij} = 0$) 
the equation (12) describes the motion of a {\it four-dimensional} fluid.
The reason for working in four-dimensions is that no 
single-energy discrete lattice exists
in three-dimensional space which ensures isotropy of the fourth order tensor
$T_{klmn} = \sum_i c_{ik}c_{il}c_{im}c_{in}$.

Three-dimensional motion is then recovered by quenching the propagation 
along the discrete speeds projecting upon the fourth dimension $x_4$. 
Quenching the fourth dimension releases
an excess of populations versus discrete speeds 
(24 against 18) introducing 'de facto' 
a degeneracy of the 'momentum eigenstates' ($c_{ik}$) 
propagating along the fourth dimension, since
each of these links carries two populations with opposite speeds along $x_4$.

Generally, to the purpose of describing sheer hydrodynamics, there's
no point of keeping two distinct populations ("doublet") 
along the same discrete speed. A 18 speed-18 populations ({\it 18s18p} for brevity)
scheme is perfectly adequate for three-dimensional Navier-Stokes equations.
If, on the other hand, doublets are retained, then the current $J_4$ is readily shown
to be passively trasported by the three-dimensional fluid; this means that 
the {\it 24s18p} model delivers the 3D Navier-Stokes equations plus a passive scalar. 
The same reasoning can be down-iterated to lower dimensions leading
to {\it 9s24p}, {\it 9s18p} (two-dimensional fluid plus one/two passive scalars respectively) 
and {\it 3s24p}, {\it 3s18p} {\it 3s9p}(one dimensional fluid and one/two/three passive scalars) in 1D.
A simple sketch illustrates the tree structure associated with dimensional quenching
of the FCHC scheme:
\newpage
\begin{verbatim}
               24             4D: 4D fluid, 0 passive scalar (24s24p)
             *    -
            18     6          3D: 3D fluid, 0 passive scalar (18s18p)
          *    -     -
         9      9     6       2D: 2D fluid, 0 passive scalar  (9s9p)
       *   -      -     -                   2                 (9s24p)
      3     6      9      6   1D: 1D fluid, 0 passive scalar  (3s3p)
                                            1                 (3s9p)
                                            2                 (3s18p)
                                            3                 (3s24p)
\end{verbatim}

Since passive/active scalars are in heavy use in
numerical combustion to describe fast chemistry regimes, we argue that
the aforementioned properties point to the FCHC Lattice Boltzmann as
to a potential candidate for the numerical description of combustion phenomena.

In this paper, we shall focus on two-dimensional reactive flow dynamics
with one passive scalar describing the fuel mixture fraction $Z$ 
and a second, active, scalar accounting for temperature dynamics.
This picks up the {\it 9s24p} model.
It should be noted that within the "mixed-is-burned" assumption, the
temperature field can be obtained directly from the knowledge of the
mixture fraction $Z$ via the piecewise linear mapping given by eq.(10-11).
We choose however to track the temperature field independently of $Z$ in
order to gain a qualitative feeling for the validity of the adiabatic approximation.

With these preparations, the reactive flow variables are defined 
as follows:

$\rho = \sum_{i=1}^{24}   f_i$, 
$\rho u = \sum_{i=1}^{24} f_i c_{i1}$, 
$\rho v = \sum_{i=1}^{24} f_i c_{i2}$,
$\rho Z = \sum_{i=1}^{24} f_i c_{i3}$,
$\rho T = \sum_{i=1}^{24} f_i c_{i4}$

The second scalar $T$ (temperature) is 'activated' by adding a source term, say $q$ to 
all populations projecting upon the fourth dimension.
A simple momentum budget fixes $q$ as a function of the heat released $ \dot Q$.
This is all we need to describe fast combustion.

\section{Numerical results}

As a test-case application, we shall consider a 
{\it coflow two-dimensional methane-air laminar flame} in a 
plane domain. The numerical set-up is as follows:

{\it Inlet}: Coflowing Fuel (methane)-Oxidizer (air) at $T=T_{in}$. 
The fuel enters the domain within an inner region
centered about the midline and width $H/8$.
The mixture fraction at the inlet is $Z_{in}$.
{\it Outlet}: Zero cross-flow and zero axial derivatives
of the streamwise velocity.
Upper and lower boundary are kept at a constant wall temperature $T_w$.
Other simulation parameters are:
inlet mixture fraction profile: $Z_{in}=2.0$, $Z_{st} G(y)$, $G$ being 
a gaussian profile of width
$H/8$, inlet temperature: $T_{in}=0.0$ 
(conventionally 800 K), inlet flow: Poiseuille profile, 
maximum $U_{in}$=0.1, aspect ratio $L/H = 4$.

Our baseline computation is performed on a $256x64$ grid and takes about 
$1 \mu s/step/site$ on a SUN Sparc1 workstation, ninety percent of which due to hydrodynamics.
Since no turbulence model has been implemented, the
flow viscosity ($\nu = 0.01$) corresponds to a Reynolds number 
$\frac{2 U_{in} H}{3 \nu} \sim 420$. The simulation is run over 3-5 axial residency times
$\tau_R=1.5 L/U_{in} = 3840$.

Chemistry is represented by the following one-step global reaction:
\begin{equation}
CH_4 + 2 O_2 = 2 H_2 0 + CO_2
\end{equation}

which is supposed to proceed instantly at the stoichiometric surface.

For practical purposes, the reaction is assumed to occurr within
a shell centered about $Z=Z_{st}=0.057$ with thickness $Z_{th}= 0.2 Z_{st}$.
The heat source is assumed a (piecewise) constant within this shell.

In Fig.1 we show the equicontours of the mixture fraction at
time t=4,000, corresponding to about one recirculation time.
From this picture the typical shape of a laminar
diffusion flame is apparent \cite{LIN,SMOOKE}. 
Note that the stoichiometric surface is
confined to the initial region of the domain ($x \simeq 50$ lattice units).
For a given set of physical-geometrical parameters, this
location, namely the flame length $L_f$, depends solely on $Z_{in}$, i.e
the degree of mixture richness.

In Fig 2 we show the mixture fraction at the midline $y=H/2 $ as a function
of the streamwise coordinate $x$ at two different time snapshots $t=2000,4000$.
The advancement of the mixture front is well visible along with
its diffusive spreading. As expected, after an initial diffusive transient lasting
about $\tau_R/Re \simeq 100$ time steps, the tip of the front proceeds on 
a quasi-convective regime at speed $U_f = 2U_{in}/3$. 
The flame however stabilizes around the stoichiometric surface long before
the tip comes to the outlet of the domain because $L_f$ is much 
shorter than the longitudinal span $L$ of the computational box.

Within the equilibrium chemistry assumption, mid-plane reactions ($y=H/2$)
should peak around the stoichiometric line $x=x_{st}$, where 
$Z(x_{st}, y=H/2) = Z_{st}$.
In our simulation, $x_{st} \simeq 52$ as highlighted by the 
horizontal line in Fig.2.
This is the region where a substantial temperature buildup is expected to occurr.
Such a prediction is indeed reflected by Fig 3, where the median temperature
profile is reported as a function of the axial coordinate $x$.
In line with the behaviour of the mixture fraction, the temperature
field shows a peak centered about the stoichiometric region, with
a top temperature of about 2100 K, slightly below the stoichiometric temperature
$T_{st}=2263 K$. 
For the sake of comparison, the adiabatic profile deduced by the theoretical
"frozen-in" mapping $T(Z)$ computed
with $Z$ at $t=4000$ is reported right above (dotted line on top).
From this figure, we see that a relatively good match is obtained in 
in the post-flame region, while in the pre-flame region the adiabatic profile 
provides a gross overestimate of the temperature field.
This is not surprising, since the near-inlet region is characterized by
strong gradients which make the adiabatic assumption rather questionable
\cite{BUCK}. 



Finite-rate chemistry effects are typically highlighted by means of 
temperature-mixture fraction scattergrams, such as the one presented in Fig. 4.
This figure, referring to the transversal section $x=40$ after $4,000$ time
steps, conveys a clear idea of the cooling effect produced by diffusive processes.

Part of these diffusive processes is likely to result from lattice discreteness
as well, and in particular from the finite-thickness $Z_{th}$ of the reaction
region.
These numerical effects can be controlled by a careful tuning of the discretized
version of the Dirac's delta function $\delta ( Z - Z_{st})$, but no systematic
studies have been conducted in this respect. 

In Fig 5 we present the transverse temperature profile at 
three different axial locations $x=1$ (inlet), $x=L_f/2$ (mid-flame) 
and $x=128$ (mid-domain).
As expected, a doubly-humped profile is observed within the 
flame extension, with the two peaks centered  
about the stoichiometric surface (compare with Figure 1).
Beyond the flame front a typical gaussian profile is recovered.
The overall structure of the temperature field is finally represented
in Fig 6, where the equicontours of the temperature field are displayed.
From this figure, the transition between double to single humped transverse
profiles is even more apparent.

In conclusion we have shown that minor modifications to the basic FCHC
lattice Boltzmann scheme permit to deal with reactive flows in the
limit of infinitely fast chemistry.
Given the fact that chemistry is intrinsically local in space and Lattice Boltzmann is
a proved parallel performer on its own even for inert flows, excellent performances
on parallel machines may be easily anticipated.
This said, it is worth emphasizing that this
work is still very preliminary in character and several extensions are
required before quantitative comparisons with state-of-the art numerical 
combustion techniques can be attempted.
Among others, important future developments are the following ones:

\begin{enumerate}
 \item{Two-dimensional laminar to three-dimensional turbulent flames}
 \item{Infinitely fast to finite-rate chemistry}
 \item{Nearly incompressible to compressible flows}
\end{enumerate}

The first item does not require {\it any} modification to the scheme presented 
here, but just higher numerical resolution.
To convey a quantitative idea of how far we can go with {\it direct} simulations, 
let us mention that a purely hydrodynamic model (18s18p) for isothermal turbulent 
channel flows is currently achieving Reynolds numbers of about $3,000$ on the massively parallel
computer Quadrics, with a sustained speed over $10$ Gflops \cite{JFR}. 
The numerical exploration of this regime should provide valuable informations 
on the morphology of turbulent flames.
Higher Reynolds numbers will call for suitable turbulence models.
These can easily be incorporated within the LBE method so long as 
simple algebraic turbomodels such as Smagorinski, Baldwin-Lomax 
and Renormalization Group models are used.
All what is needed is to let the scattering matrix $A_{ij}$ a local
function of the stress tensor, i.e. a pointwise function of the populations
\cite{SOM}.

Moving beyond the fast chemistry regime requires
the development of a suitable multispecies transport scheme able to
handle several scalars in three dimensions.
Since in the conventional Lattice Boltzmann formalism
each extra-scalar requires six moving populations, a naive extension of the
present scheme does not look very appealing.
However it seems possible to devise modified LBE schemes which are
able to track passive scalars using only population per species
\cite{HUDONG}.
An alternative option is to couple the present LBE scheme to 'conventional'
grid-based methods.



At variance with the previous two, the third item
{\it does} involve substantial changes to the scheme proposed in this work.

The scheme presented in this paper belongs to the simplest class of combustion
models; those in which density is not able to keep up dynamically
with significant temperature excursions.
So, in principle, the method should only be applied to 'cold' flames with little
heat release in which the main focus is on flame morphology (i.e. the scenario 
pertaining to item 1. in the aforementioned list of developments).

The next class of problems to be adressed is low-Mach reacting flows in which
density is allowed to respond to temperature changes over a significant
dynamical range of values.
The possibility to model this class of flows by means of appropriate phenomenological
extensions of the present LBE scheme is currently under investigation.

Finally, one should consider the class of fully-compressible, high-Mach flows in
which density is supposed to respond to both temperature and pressure changes.
We believe such class is definitely out of reach for the present LBE model.

In fact, in order to capture full thermo-hydrodynamic behaviour the discrete speeds
must necessarily accomodate multienergy levels.
A few models in this class  have already been proposed
\cite{CHE2,YCH}, but they still need to be perfected in many respects.
First, they involve many discrete speeds ($40$ in 3D), let alone the additional
degrees of freedom explicitly required to incorporate passive/active scalar dynamics; 
second, and more importantly, it appears like their stability is still matter of 
concern \cite{ALD}.
Further research work is definitely required to clarify whether discrete speed
models (of reasonable complexity) will ever be able to perform competitively for
full thermo-hydrodynamic combustion applications.

Meanwhile, we believe that the model presented
here, although clearly limited in scope, can provide a useful starting point
to explore phenomenological alternatives to fully-fledged discrete speed models
for the numerical simulation of (some class of) combustion phenomena.

\newpage
\section{Figure captions}

\begin{itemize}
 \item Fig.1: Isocontours of the mixture fraction field $Z(x,y)$ after $4,000$ time steps.
              The stoichiometric line corresponds to $Z=0.057$.

 \item Fig.2: The axial profile of the mixture fraction at the midline $y=128$, after
              $2,000$ and $4,000$ time steps. The solid line is associated with the
              stoichiometric value $Z_{st} = 0.057$.

 \item Fig.3: The axial profile of the temperature field at the midline $y=128$, after
              $2,000$ and $4,000$ time steps. The solid line is the initial profile
              and the curve labeled 'Tadiabatic' represents the adiabatic profile
              computed with the mixture field $Z$ at $t=4,000$.

 \item Fig.4: Temperature versus mixture fraction scatterplot at $x=40$
              at time $t=4,000$.. 

 \item Fig.5: The transverse temperature profile at three axial locations, $x=1$ (inlet),
              $x=25$ (mid-flame) and $x=128$ (mid-domain) at $t=4,000$.. 

 \item Fig.6: Isocontours of the temperature field after $4,000$ time steps.

\end{itemize}
\end{document}